\documentclass[conference]{IEEEtran}
\IEEEoverridecommandlockouts
\usepackage{tikz}
\usepackage{graphicx}
\PassOptionsToPackage{hyphens}{url}\usepackage{hyperref}

\usepackage{ctable}
\usepackage{fancyvrb}

\usepackage{multirow} 
\usepackage{float}
\usepackage{pifont}
\usepackage{array}
\usepackage{listings}
\usepackage{algorithm}
\usepackage[noend]{algpseudocode}
\usepackage{fancyhdr}
\usepackage{fvextra}
\usepackage{xcolor}
\usepackage[utf8]{inputenc}
\usepackage{orcidlink}
\usepackage{hanging}
\def\BibTeX{{\rm B\kern-.05em{\sc i\kern-.025em b}\kern-.08em
    T\kern-.1667em\lower.7ex\hbox{E}\kern-.125emX}}

\DeclareUnicodeCharacter{00A0}{ }

\DeclareUnicodeCharacter{2514}{\mbox{\kern.23em\vrule height2.2exdepth-1.8ptwidth.4pt\vrule height2.2ptdepth-1.8ptwidth.23em}}
\DeclareUnicodeCharacter{2500}{\mbox{\vrule height2.2ptdepth-1.8ptwidth.5em}}

\DeclareUnicodeCharacter{251C}{\mbox{\kern.23em
  \vrule height2.2exdepth1exwidth.4pt\vrule height2.2ptdepth-1.8ptwidth.23em}}

\makeatletter
\renewcommand{\ALG@name}{Code}
\makeatother

\definecolor{codegreen}{rgb}{0,0.6,0}
\definecolor{codegray}{rgb}{0.9,0.9,0.9}
\definecolor{makeblue}{rgb}{0.9,0.9,1.0}
\definecolor{codegray2}{rgb}{0.2,0.2,0.2}
\definecolor{codepurple}{rgb}{0.58,0,0.82}
\definecolor{backcolour}{rgb}{0.95,0.95,0.92}

\lstdefinelanguage{Fortran2023}
{
  keywords = {parallel, loop, if, in, while, do, concurrent, else,  target,enddo, end},
  comment = [l]{//},
}

\lstdefinelanguage{Compile}
{
  keywords = {},
  comment = [l]{//},
}

\lstdefinestyle{mystyle}{
    backgroundcolor=\color{codegray},   
    commentstyle=\color{codegreen},
    keywordstyle=\color{codepurple},
    stringstyle=\color{codepurple},
    numberstyle=\tiny\color{codegray2},
    basicstyle=\ttfamily\footnotesize,
    breakatwhitespace=false,         
    breaklines=true,                 
    captionpos=b,                    
    keepspaces=true,                                
    showspaces=false,                
    showstringspaces=false,
    showtabs=false,                  
    tabsize=2
}
\lstdefinestyle{mystyle2}{
    backgroundcolor=\color{makeblue},   
    basicstyle=\ttfamily\footnotesize,
    breakatwhitespace=false,         
    breaklines=true,                 
    captionpos=b,                    
    keepspaces=true,                                
    showspaces=false,                
    showstringspaces=false,
    showtabs=false,                  
    tabsize=2
}

\graphicspath{{./}{figures/}}

\begin{document}

\title{Portability of Fortran's `do concurrent' on GPUs}

\author{
\IEEEauthorblockN{Ronald M. Caplan \orcidlink{0000-0002-2633-4290}}
\IEEEauthorblockA{\textit{Predictive Science Inc.}\\
San Diego, CA USA\\
0000-0002-2633-4290}
\and
\IEEEauthorblockN{Miko M. Stulajter \orcidlink{0000-0003-0939-1055}}
\IEEEauthorblockA{\textit{Predictive Science Inc.}\\
San Diego, CA USA\\
0000-0003-0939-1055}
\and
\IEEEauthorblockN{Jon A. Linker \orcidlink{0000-0003-1662-3328}}
\IEEEauthorblockA{\textit{Predictive Science Inc.}\\
San Diego, CA USA\\
0000-0003-1662-3328}
\and
\IEEEauthorblockN{Jeff Larkin \orcidlink{0000-0001-7132-3120}}
\IEEEauthorblockA{\textit{NVIDIA Corporation}\\
Santa Clara, CA USA\\
000-0001-7132-3120}
\and
\IEEEauthorblockN{Henry A. Gabb \orcidlink{0000-0002-9507-4250}}
\IEEEauthorblockA{\textit{Intel Corporation}\\
Austin, TX USA \\
0000-0002-9507-4250}
\and
\IEEEauthorblockN{Shiquan Su \orcidlink{0000-0002-9546-6449}}
\IEEEauthorblockA{\textit{Intel Corporation}\\
Boulder, CO USA \\
0000-0002-9546-6449}
\and
\IEEEauthorblockN{Ivan Rodriguez}
\IEEEauthorblockA{\textit{Hewlett Packard Enterprise}\\
Austin, TX USA \\
ivan.rodriguez@hpe.com}
\and
\IEEEauthorblockN{Zachary Tschirhart \orcidlink{0000-0002-2968-6531}}
\IEEEauthorblockA{\textit{Hewlett Packard Enterprise}\\
Austin, TX USA \\
0000-0002-2968-6531}
\and
\IEEEauthorblockN{Nicholas Malaya \orcidlink{0000-0001-6259-7453}}
\IEEEauthorblockA{\textit{Advanced Micro Devices Inc}\\
Austin, TX USA \\
0000-0001-6259-7453}
}

\maketitle

\begin{abstract}
There is a continuing interest in using standard language constructs for accelerated computing in order to avoid (sometimes vendor-specific) external APIs.  For Fortran codes, the {\tt do concurrent} (DC) loop has been successfully demonstrated on the NVIDIA platform.  However, support for DC on other platforms has taken longer to implement.  Recently, Intel has added DC GPU offload support to its compiler, as has HPE for AMD GPUs.  In this paper, we explore the current portability of using DC across GPU vendors using the in-production solar surface flux evolution code, HipFT.  We discuss implementation and compilation details, including when/where using directive APIs for data movement is needed/desired compared to using a unified memory system.  The performance achieved on both data center and consumer platforms is shown.
\end{abstract}

\begin{IEEEkeywords}
Fortran, accelerated computing, OpenMP, OpenACC, do concurrent, standard language parallelism.
\end{IEEEkeywords}

\section{Introduction}
\label{sec:intro}

The use of standard language parallelism for accelerated computing has been gaining popularity in the last few years.  This includes C++ parallel algorithms \cite{cplusplusstdparnasa}, Fortran's {\tt do concurrent} (DC) \cite{stdpar_dc_waccpd}, and drop-in replacements for Python's {\tt numpy} such as {\tt cunumeric} \footnote{\url{https://developer.nvidia.com/cunumeric}}.  These new language features have the potential to eliminate (or greatly reduce) the need for external APIs, leading to better portability \cite{Wolfe2021,compatchart} and persistence.  

Our focus in this paper is on Fortran codes using DC.  The {\tt do concurrent} construct is an alternative to Fortran's {\tt do} loop which can be used for loops that have no data dependencies (and can therefore be computed out-of-order\footnote{\url{https://flang.llvm.org/docs/DoConcurrent.html}}).  Since this often implies the loop can be executed in parallel, the DC loop can be used by compilers as a hint that the loop can be parallelized.  The addition of locality specifiers (such as {\tt local}, {\tt shared}, etc.) and the new {\tt reduce} clause in the Fortran 2023 specification gives developers additional ways to express details of the loop's ability to be parallelized.  The parallelization of a DC loop can be performed on multicore CPUs as well as offloaded to GPU accelerators.  The portability of DC parallelization for CPUs across compilers/vendors was explored in \cite{stdpar_dc_waccpd} and \cite{BabelStream:Fortran}, where it was found that the DC implementations produced similar results to using classic OpenMP \cite{openmp_book} \footnote{The OpenMP name is a registered trademark of the OpenMP Architecture Review Board} multi-code parallel directives.  

In this paper, we focus exclusively on GPU offload with DC. While there have been previous promising results on the NVIDIA platform (see the next section), how well they extend to other vendors is an open question.  Here, we describe testing a medium-sized production code on NVIDIA, Intel, and AMD GPUs.  We test the current status of vendor-compiler combinations in being able to offload DC to GPUs and explore using both a `pure Fortran' approach versus using minimal OpenMP and/or OpenACC \cite{openacc_book} directives (for CPU-GPU data management and/or to augment the DC implementations when needed/wanted for optimization).

The paper is organized as follows.  Sec.~\ref{sec:relwork} describes related work on using Fortran's DC for accelerated computing.  In Sec.~\ref{sec:hipft}, we describe the production code HipFT and its DC implementation.  How the code is built and run across vendor platforms is described in Sec.~\ref{sec:build}.  The simulation used to test the implementations is described in Sec.~\ref{sec:test}, with the results for both server/data center and consumer hardware shown in Sec.~\ref{sec:results}.  We summarize the current status and future for the portability of Fortran standard parallelism in Sec.~\ref{sec:summary}.

\section{Related Work}
\label{sec:relwork}

The first compiler to support DC for accelerated computing was NVIDIA's HPC SDK in November of 2020 for NVIDIA GPUs \footnote{\url{https://developer.nvidia.com/blog}\\ \url{accelerating-fortran-do-concurrent-with-gpus-and-the-nvidia-hpc-sdk}}, followed by the Intel IFX compiler in 2022 for Intel GPUs\footnote{\url{https://www.intel.com/content/www/us/en/developer/articles/technical/using-fortran-do-current-for-accelerator-offload.html}}, and most recently, the HPE CCE compiler for some NVIDIA and AMD GPUs\footnote{\url{https://cpe.ext.hpe.com/docs/cce}}.

DC support for GPUs is quite new but has already seen adoption in benchmarks, mini apps (such as miniWeather\footnote{\url{https://github.com/mrnorman/miniWeather/tree/main/fortran}} included in the SPEChpc2021 benchmark suite\footnote{\url{https://www.spec.org/hpc2021}}), small tools, and some large production codes.  

In \cite{BabelStream:Fortran}, the BabelStream benchmark was tested using only DC on both CPUs and GPUs.  The authors found that the performance on CPUs and GPUs was similar to the implementations using external APIs such as OpenACC, OpenMP, and CUDA.  

A similar result was shown for the tool `diffuse' in \cite{stdpar_dc_waccpd}, where the GPU results were similar to using DC alone compared to DC with OpenACC for manual CPU-GPU data management.  

Conversely, for the production tool `POT3D' \cite{caplan2021variations} (also part of SPEChpc2021, consisting of a preconditioned conjugate gradient solver), it was found that, due to GPU-aware MPI calls, manually managing the data with a few OpenACC directives was faster than using the NVIDIA automatic managed memory, especially across multiple discrete GPUs \footnote{\url{https://www.nvidia.com/en-us/on-demand/session/gtcspring22-s41318}}.  This performance difference was shown to be eliminated when using the newer unified memory features of the NVIDIA Grace-Hopper (GH) architecture (at least for a single GH super chip)\footnote{\url{https://developer.nvidia.com/blog/simplifying-gpu-programming-for-hpc-with-the-nvidia-grace-hopper-superchip}}.  

The most computationally expensive kernels of the quantum chemistry production code GAMESS were tested using DC alone on CPUs and GPUs in \cite{gamess_dc}. The authors discovered that running the DC implementation on a single GPU surpassed the performance of their earlier OpenACC and OpenMP target implementations.  This work was recently extended to the LibERI library interfaced to GAMESS, where the performance of the DC code outperformed their OpenMP Target implementation \cite{liberi_gamess_dc}.

Another production code that has implemented DC for GPU offload is the solar physics code MAS, as described in \cite{caplan2023acceleration}. Similar to POT3D, the authors observed a performance drop when using DC alone compared to the OpenACC implementation due to GPU-aware MPI calls. However, most of the lost performance was recovered by incorporating a small number of OpenACC directives for manual data management.

DC was also used to GPU-accelerate the production relativistic MHD astrophysics code `ECHO' \cite{del2024gpu}. The authors relied on automatic managed memory, but still had to use some OpenACC directives for the code to work correctly (mostly dealing with subroutines within DC loops). They demonstrated impressive scaling up to 256 quad-GPU nodes in cases where the computation time constituted a much larger fraction of the run time compared to MPI communication.

While all these implementations are promising (and many yielded on-par performance with other acceleration methods), they all exclusively used the NVIDIA platform for GPUs.  In this study, we investigate the portability of DC by additionally testing it on the Intel and AMD platforms.  

\section{High Performance Flux Transport (HipFT)}
\label{sec:hipft}

The code we use to test the portability of DC for GPU offload is the High Performance Flux Transport (HipFT) code (\url{github.com/predsci/hipft}).  It is the computational core of the Open-source Flux Transport (OFT) model (\url{github.com/predsci/oft}) which simulates the evolution of the radial magnetic field on the surface of the Sun \cite{Pogorelov_2024}. Full-Sun maps of this field play a critical role in solar and heliospheric physics research, but the fields are often measured only along the Sun-Earth line. Simulating surface evolution helps mitigate this data gap and can also be used independently to study solar surface dynamics, the solar dynamo, and the solar cycle \cite{Yeates2023}.

HipFT implements advection, diffusion, and data assimilation and can compute multiple realizations in a single run across model parameters to create ensembles of full-Sun maps. It employs finite difference methods on a spherical grid and utilizes highly accurate and high-performance numerical methods, such as a Strong Stability Preserving Runge-Kutta (SSPRK(4,3) \cite{ssprk_book}) scheme with a Weighted Essentially Oscillatory (WENO3-CS(h) \cite{WENO3CS}) scheme for advection, and the Extended Stability Runge-Kutta Gegenbauer Super Time Stepping (RKG2(3/2) \cite{RKG2b}) scheme for diffusion.  Computationally, these methods boil down to memory-bound array and stencil operations.   HipFT is written in modern Fortran ($\sim 9000$ lines of code) and parallelized for use with multiple multi-core CPU and multi-GPU compute nodes.  Details of the code are described in Ref.~\cite{oft1}.

\subsection{Parallel implementation}
\label{sec:hipft_dc}

HipFT was developed as a DC-based code from the ground up. An example DC loop is shown in Listing \ref{lst:dc}.  To parallelize across multiple multi-GPU and/or multi-CPU nodes, we use a combination of DC, Open(MP/ACC) directives, and MPI.  
\begin{lstlisting}[language=Fortran2023, caption={Example DC loop in the HipFT code.  It computes the diffusion operator matrix-vector product for internal grid points.}, label={lst:dc}]
do concurrent (i=1:nr,k=2:npm-1,j=2:ntm-1)
  y(j,k,i) =  coef(j,k,1,i)*x(j,  k-1,i)  & 
            + coef(j,k,2,i)*x(j-1,k  ,i)  &
            + coef(j,k,3,i)*x(j  ,k  ,i)  &
            + coef(j,k,4,i)*x(j+1,k  ,i)  &
            + coef(j,k,5,i)*x(j,  k+1,i)
enddo
\end{lstlisting}
MPI is used to evenly spread the realizations of maps across compute devices (either GPUs or CPUs).  Each realization is an independent calculation so there are no peer-to-peer MPI communications needed.  However, in order to run all of them together, the strictest numerical time step limit of all realizations is used by all MPI ranks (requiring collective scalar operations).  Other collective operations are used in the analysis step of the simulation, where maximum, minimum, and integral quantities of each map are collected and written to a text file.  When the maps need to be output, they are first transferred from the GPUs to the CPU, and then combined using MPI before being written to disk.

For scalar reduction loops, we make use of the Fortran 2023 DC {\tt reduce} clause. However, since this is a recent addition to Fortran, many compilers do not yet recognize it, including GCC.  Therefore, when using GCC on the CPU, we pre-process the code to remove the {\tt reduce} clauses. This does not affect the correctness of the code because GCC does not directly support multithreading with DC. Instead, we utilize GCC's CPU auto parallelization feature for the DC loops, which auto-detects the reductions (see Ref.~\cite{stdpar_dc_waccpd} for details).  

An often difficult algorithm to parallelize on GPUs are array reductions (see \cite{caplan2019gpu}).  Depending on the order of the loops, either an atomic operation is needed for each reduction array element (which may not be efficient), or several scalar reductions are required, which may be serialized unless the topology of the GPU parallelism is optimally chosen by the compiler (such as spreading the array across compute blocks, and using threaded parallelism within each block with shared memory).  In HipFT, the array reductions use the second approach.  An example of an array reduction within the code is shown in Listing \ref{lst:arrayreduce}.
\begin{lstlisting}[language=Fortran2023, caption={Example DC array reduction loop in the HipFT code.  It computes the polar boundary conditon for the diffusion operator.}, label={lst:arrayreduce}]
do concurrent(i=1:nr)
  fn = zero
  fs = zero
  do concurrent(k=2:npm-1) reduce(+:fn,fs)
    fn = fn + (diffusion_coef(1    ,k,i)        &
             + diffusion_coef(2    ,k,i))       &
            * (x(2  ,k,i) - x(1    ,k,i))*dp(k)
    fs = fs + (diffusion_coef(nt-1 ,k,i)        &
             + diffusion_coef(nt   ,k,i))       &
            * (x(ntm,k,i) - x(ntm-1,k,i))*dp(k)
  enddo
  do concurrent(k=1:npm)
    y(  1,k,i) =  fn*bc_diffusion_npole_fac
    y(ntm,k,i) = -fs*bc_diffusion_spole_fac
  enddo
enddo
\end{lstlisting}

A key issue to address when running on GPU accelerators is data management.  Nearly all GPUs have their own separate memory which may or may not be directly connected to the host CPU.  Often, the time required to transfer data to and from the GPU is extremely slow, making it crucial to keep the computation within GPU memory as much as possible.  The Fortran language currently does not have any concept of distinct memory spaces.  Therefore, in order to run pure DC codes efficiently, an automatic data management system must exist in the compiler and/or hardware.  In the absence of such automatic memory management, managing the data manually with external APIs is required for good performance.  To maximize portability and performance, HipFT uses OpenMP target directives for manual data movement.  Although the developers of HipFT prefer OpenACC, OpenMP's greater portability across vendors outweigh this preference.

Another important issue comes up when running on multiple GPUs using MPI.  In this case, each MPI rank needs to indicate which GPU device to run on.  Fortran has no concept of multiple compute devices in the language. Therefore, we once again rely on external APIs to select the device (although in some cases, using an alternative launching method can avoid this requirement \cite{caplan2023acceleration}).  HipFT uses both a single OpenACC and an OpenMP target API directive (shown in Listing \ref{lst:device}) to set the device number (both directives are necessary for the NVIDIA compiler due to their use of an OpenACC backend for DC, independent of the OpenMP target backend).
\begin{lstlisting}[language=Fortran2023, caption={Directives used to set the GPU device number in HipFT assuming 1 GPU per MPI rank, where {\tt iprocsh} is the MPI shared rank.}, label={lst:device}]
!$ call omp_set_default_device (iprocsh)
!$acc set device_num(iprocsh)
\end{lstlisting}
 
\section{Building HipFT across vendors}
\label{sec:build}

\lstset{style=mystyle2,language=Compile,float,floatplacement=htbp}

Here we describe how we build and run HipFT across GPU vendors, including any modifications needed for optimizing performance.   While server/data center GPUs are the focus here, we also discuss building on consumer GPUs, as they are useful for small-to-medium simulations and development.  We cover compilers from NVIDIA's HPC SDK Toolkit, Intel's OneAPI HPC Toolkit, and HPE's Cray Compiler Environment.  However, we note that the {\tt Flang} compiler developers are working to support DC GPU offload, which will make it an alternative to use in the future.  GCC may also eventually support DC GPU offload, but currently does not support direct parallelization of DC at all (for details on building and running DC codes like HipFT in parallel on CPUs using GCC, see Ref.~\cite{stdpar_dc_waccpd}).

\subsection{NVIDIA}
\label{sec:build_nvidia}

NVIDIA has the most mature implementation of DC for GPU offload.  The NV HPC SDK compiler {\tt nvfortran} (NV) version 24.5 is used to compile the code.  Under the hood, the compiler uses its OpenACC backend to compile the DC loops.  As discussed above, this requires us to add an OpenACC device selection directive in addition to the equivalent OpenMP directive.

NV has three memory management options: {\tt mem:separate}, {\tt mem:managed}, and {\tt mem:unified}.  {\tt mem:separate} tells the compiler not to automatically manage the GPU-CPU memory (i.e., it allows the user to manually manage the memory using external APIs). 

{\tt mem:managed} activates  automatic memory management for non-static arrays, which under the hood translates to using {\tt cudaMallocManaged()} for all allocations.  This managed memory works for all NVIDIA GPUs, including discrete GPUs.  It has been shown to be a productive means to get good results \cite{BabelStream:Fortran,del2024gpu}, and can be similar in performance to manually managed memory \cite{stdpar_dc_waccpd}.  However, sometimes (especially when using CUDA-aware MPI) the managed memory can result in a reduction of performance \cite{caplan2023acceleration}.
{\tt mem:unified} is the newest memory management option and is available on unified CPU-GPU architectures such as the GH superchip, and on systems using the open source NVIDIA kernel driver which supports heterogeneous managed memory (HMM).  In this mode, all memory (even static) gets auto-paged between GPU and CPU memory in a unified addressed space, and, at least for a single GH, has been shown to perform the same (or better) than manual memory management. If a memory mode is not specified via a compiler flag, {\tt unified} mode is used by default when building on a machine with support for unified memory or {\tt managed} mode otherwise. 

When using {\tt separate} mode it is necessary (for good performance) for the developer to ensure that data is accessible on the GPU, usually through OpenACC or OpenMP data directives. These directives are not necessary in {\tt managed} or {\tt unified} modes, but if they are present, the NVIDIA compiler will use them to provide hints to the driver about the usage patterns of the data. These hints can be used to tell the driver the preferred location of the data and even to prefetch the data into or out of GPU memory.

We compile HipFT with the additional flags (other than standard MPI, HDF5, optimization, and library flags):
\begin{lstlisting}[language=Compile, label={lst:nv_c1}]
-march=native -Minfo=accel -stdpar=gpu -acc=gpu
\end{lstlisting}
where we add
\begin{lstlisting}[language=Compile, label={lst:nv_c2}]
-mp=gpu -gpu=ccnative,mem:separate 
\end{lstlisting}
for manual memory management, 
\begin{lstlisting}[language=Compile, label={lst:nv_c3}]
-gpu=ccnative,mem:managed
\end{lstlisting}
for managed memory, and
\begin{lstlisting}[language=Compile, label={lst:nv_c4}]
-gpu=ccnative,mem:unified
\end{lstlisting}
for unified memory (no OpenMP is needed, so {\tt -mp=gpu} which activates OpenMP target is not used).  The flag {\tt ccnative} tells the compiler to compile for the best GPU it can find on the system, {\tt -Minfo=accel} outputs details of the compiler's offloading, {\tt -acc=gpu} activates OpenACC (needed for device selection), and {\tt -stdpar=gpu} activates the standard parallel (DC) offload.  For CPU runs, one can use {\tt -stdpar=multicore} instead of the GPU flags.

NVIDIA produces many kinds of GPUs, which can be summarized into three categories: Server/Data center, Professional graphics, and Consumer graphics.   While the two graphics categories are not designed with a focus on classic HPC applications (e.g., they have greatly reduced numbers of double precision floating point (FP) units), they are still quite powerful and usable for such applications (especially for memory-bound algorithms like those in HipFT).  Therefore, we also test HipFT on consumer cards, where the same compiler flags as the server GPUs shown above are used.

\subsection{INTEL}
\label{sec:build_intel}

Support for DC GPU offload is a recent addition (introduced in version {\tt 2022.0}) to the Intel compiler {\tt ifx} and is implemented using an OpenMP target backend.  The compiler does not currently have an automatic memory management system for DC, so to avoid the significant performance reduction from GPU-CPU transfers, we manually manage the memory using pairs of OpenMP target enter/exit data directives. These pairs form unstructured data regions, which allows multiple data entry and exit points across different regions of the code. In the original Intel implementation, the DC loops were transformed into OpenMP target loop directives with an `always' map-type-modifier indicating to always transfer the data to and from the GPU.  In the latest release (as of this writing, {\tt 2024.2}), a special {\tt maptype} compiler flag (shown in Listing \ref{lst:intel_c1})  has been added to let the programmer specify the map-type-modifier for all transformed OpenMP target regions. Setting this value to "present" tells the compiler to assume all data (other than scalars) in DC loops are already on the device, leaving the memory management for the user to take care of using OpenMP target data directives.

We compile HipFT with the additional flags (other than standard MPI, HDF5, optimization, and library flags):
\begin{lstlisting}[language=Compile, label={lst:intel_c1}]
-xHost -fp-model precise -heap-arrays 
-fopenmp-target-do-concurrent -fiopenmp
-fopenmp-targets=spir64 
-fopenmp-do-concurrent-maptype-modifier=present
\end{lstlisting}
where  {\tt -fiopenmp} activates OpenMP, \newline
{\tt -fopenmp-target-do-concurrent} activates DC offload,
{\tt -fopenmp-targets=spir64} targets the Intel GPUs, and 
{\tt -fopenmp-do-concurrent-maptype}\newline
{\tt -modifier=present} tells the compiler to assume the arrays used in DC loops are present on the GPU.  For CPU runs, we use the same compiler flags, but with {\tt spir64\_x86} instead of {\tt spir64}, and without the {\tt maptype} flag. 

In our preliminary GPU tests, we found that a few specific loops were running far slower than expected.  These were the array reductions with nested DC loops as shown in Sec.~\ref{sec:hipft_dc}.  After investigation, it was found that adding {\tt !\$omp parallel loop} above the inner DC loops changes the internal heuristics of the compiler, resulting in a far more efficient kernel (see Sec.\ref{sec:results}).  Adding these directives is not standard in a DC loop (and can cause an 
 error when using other compilers).  However, the modified heuristics are being integrated into the default compilation, so future compiler releases will not require these added directives. We indicate this small (6 lines) modification of the code in the results with an asterisk (*).

Intel's latest release of discrete consumer GPUs is the `Arc' series.  These GPUs, like NVIDIA's consumer GPUs,  are targeted for graphics applications but they can still be used for computations, making them useful for development and small-to-medium workloads.  One large caveat is Arc's lack of any double precision FP units.  In order to run double precision code, we need to activate a double precision emulation mode by setting the following environment variables:
\begin{lstlisting}[language=Compile, label={lst:intel_c2}]
IGC_EnableDPEmulation=1
SYCL_DEVICE_WHITE_LIST=""
OverrideDefaultFP64Settings=1
\end{lstlisting}
and adding the compiler flag:
\begin{lstlisting}[language=Compile, label={lst:intel_c3}]
-Xopenmp-target-backend "-device arc"
\end{lstlisting}
As using double precision emulation can be quite slow, we focus our results using Arc GPUs on correctness and value for development, not performance.  

\subsection{AMD}
\label{sec:build_amd}

Support for DC offload to AMD GPUs is recent, with the only current implementation being provided by Hewlett Packard Enterprise (HPE) through their Cray Compiler Environment (CCE).  The latest CCE compiler supports most of the DC clauses, but only a subset of scalar reductions with the {\tt reduce} clause ({\tt min} and {\tt max} are not yet supported).  This support should be in the compiler by the time of publication, as it is already included in an in-development version which is used here.

CCE currently supports DC AMD GPU offload for unified memory architectures exclusively  (limiting support to CDNA2 and CDNA3 GPUs such as the AMD Instinct\textsuperscript{TM} MI250X GPU and MI300A APU).  The MI250X is currently deployed in the Frontier machine at ORNL, while the MI300A will be featured in the upcoming El Capitan system at LLNL.

We compile HipFT with the additional flags (other than standard MPI, HDF5, optimization, and library flags):
\begin{lstlisting}[language=Compile, label={lst:amd_c1}]
-homp -hacc -hoffload_do_concurrent 
-hcpu=x86-rome -haccel=amdgcn-gfx942 
\end{lstlisting}
where {\tt -homp} and {\tt -hacc} activates the OpenMP and OpenACC runtimes respectively (both are necessary as CCE uses an OpenACC backend for DC), {\tt -hcpu} and {\tt -haccel} set the CPU and GPU hardware targets respectively, and {\tt -hoffload\_do\_concurrent} activates the DC GPU offload.  It is also necessary to set the environment variable {\tt HSA\_XNACK=1} to activate the automatic unified memory management\footnote{\url{https://rocm.docs.amd.com/en/latest/conceptual/gpu-memory.html}}.

\section{Test case}
\label{sec:test}

For each compiler/GPU/CPU tested, we first ensure correct code execution by making sure the testsuite included with the HipFT code passes.  The testsuite validation criteria uses various derived quantities of the solution which are checked to be within a relative tolerance of $1\times10^{-5}$ compared to a reference solution.

\begin{figure}[tb]
\centering
\includegraphics[width=0.475\textwidth]{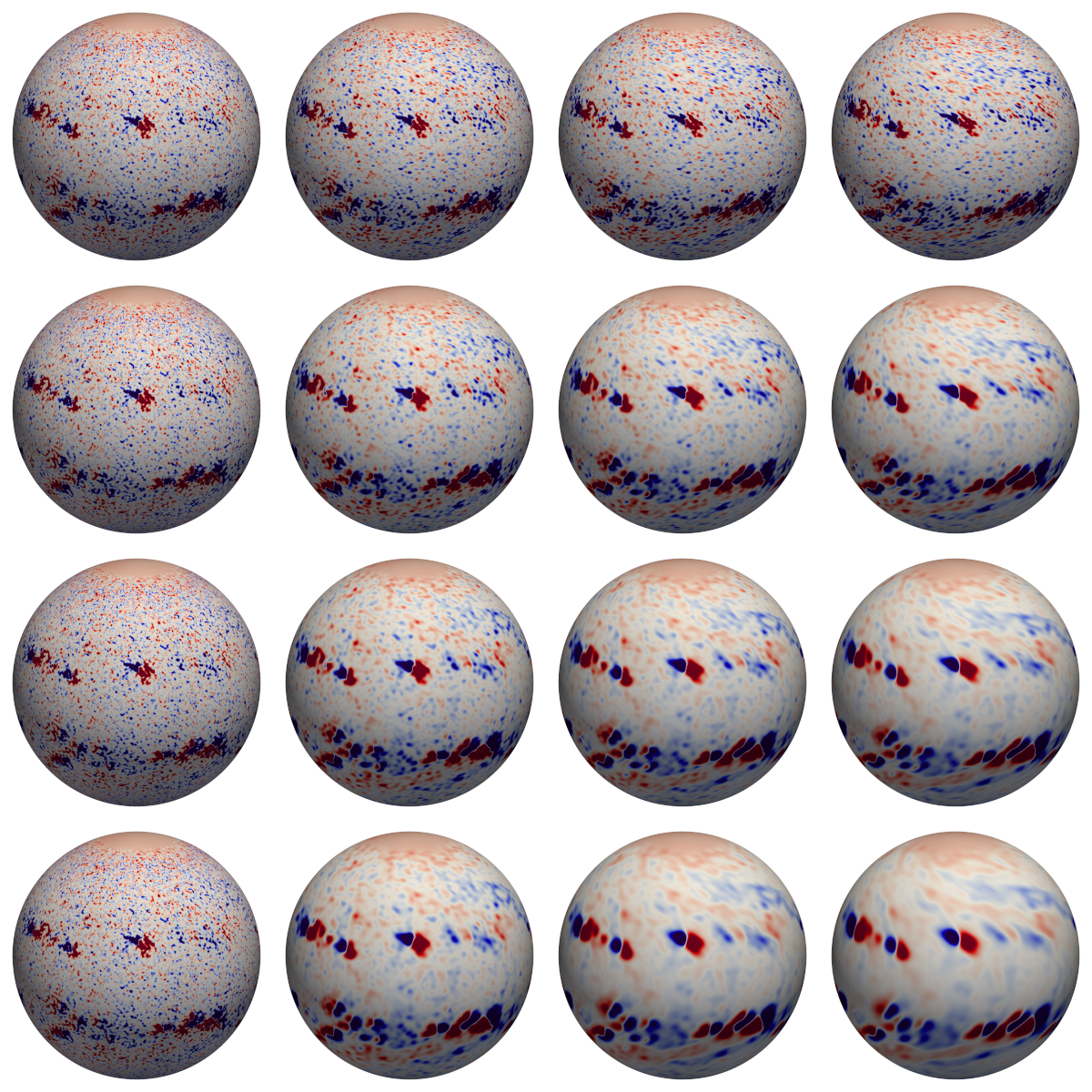}
\caption{Visualization of the HipFT test case used to evaluate the DC implementations.  We show the maps at times 0, 200, 400, and 600 hours (left to right) for four of the eight realizations (top to bottom). 
\label{fig:testrun}} 
\end{figure}

For the main results presented here, we use the included example run {\tt flux\_transport\_1rot\_flowAa\_diff\_r8} as our test case.  This run starts with a pole-filled, processed full-Sun `synoptic' map for Carrington rotation 2252 at a resolution of 1024x512.  It simulates surface flux transport (using both advection and diffusion) for 28 days (approximately one solar rotation) using analytically prescribed meridional and differential rotation flows.  Eight realizations are computed (allowing up to 8 MPI ranks) where four levels of diffusion are used, and a magnetic field based flow attenuation is or is not applied, leading to the eight total realizations. A visualization of the run is shown in Fig.~\ref{fig:testrun}.

The HipFT code, like many finite-difference codes, is highly performance bound by memory bandwidth.  Consequently, the relative performance across different hardware solutions is generally expected to correlate with their peak memory bandwidths.  However, while the test case used is a more realistic problem size for HipFT than the testsuite runs, it is still somewhat small from an HPC perspective.  With only 1.43 GB of memory usage, saturating the performance of large GPUs is challenging.  Given this (along with the very recent nature of DC support for Intel and AMD) the timings results shown in Sec.~\ref{sec:results} should not be taken as benchmarks of the respective vendors' hardware.  Instead, we aim to achieve reasonable/acceptable performance across all vendors with the same code.  

As mentioned in Sec.~\ref{sec:build}, HipFT uses OpenMP target directives for manually managing the GPU memory.  For NVIDIA, using these directives is optional, so we run the test case with manual memory, managed memory, and unified memory where available.  We note that when using managed or unified memory, the code becomes pure standard Fortran (with the exception of a single device selection directive).  For Intel GPUs, we must use the manual data directives for efficient computation (as unified memory is not yet supported), while for AMD GPUs with CCE, we are required to rely on automatic memory management (as directive-based data management is not currently supported by CCE).

\section{Results}
\label{sec:results}

Here we show the results of running the test case of Sec.~\ref{sec:test} across GPU vendors for both server and consumer GPUs.  We also show a couple of single node CPU results for comparison.  As mentioned above, the timing results shown here should not be taken as benchmarks of the respective vendors' hardware, as the problem size is small and DC support is quite new.

As HipFT was developed as a DC-based code from the ground up, we do not compare the performance of individual DC loops to alternative GPU-offload methods (such as OpenACC/MP, CUDA-Fortran, etc.).  For examples of such  comparisons, see Ref.~\cite{BabelStream:Fortran} and \cite{caplan2023acceleration}, where the DC performance was found to be similar to other forms of GPU-offload.  Our focus here is the portability of DC GPU offload.

Since various algorithms can perform differently across platforms, and since some systems may have slower file systems effecting input/output times, we display the timings broken into the main algorithmic components.  The final component shown in gray and labeled "other" includes MPI overheads and file input and output.  As this component can vary widely across systems and is outside the main computational parts of the code, its importance in analyzing the results is minimal.

\subsection{Data center / Server}
\label{sec:results_server}

\begin{figure}[htb]
\centering
\includegraphics[width=0.475\textwidth]{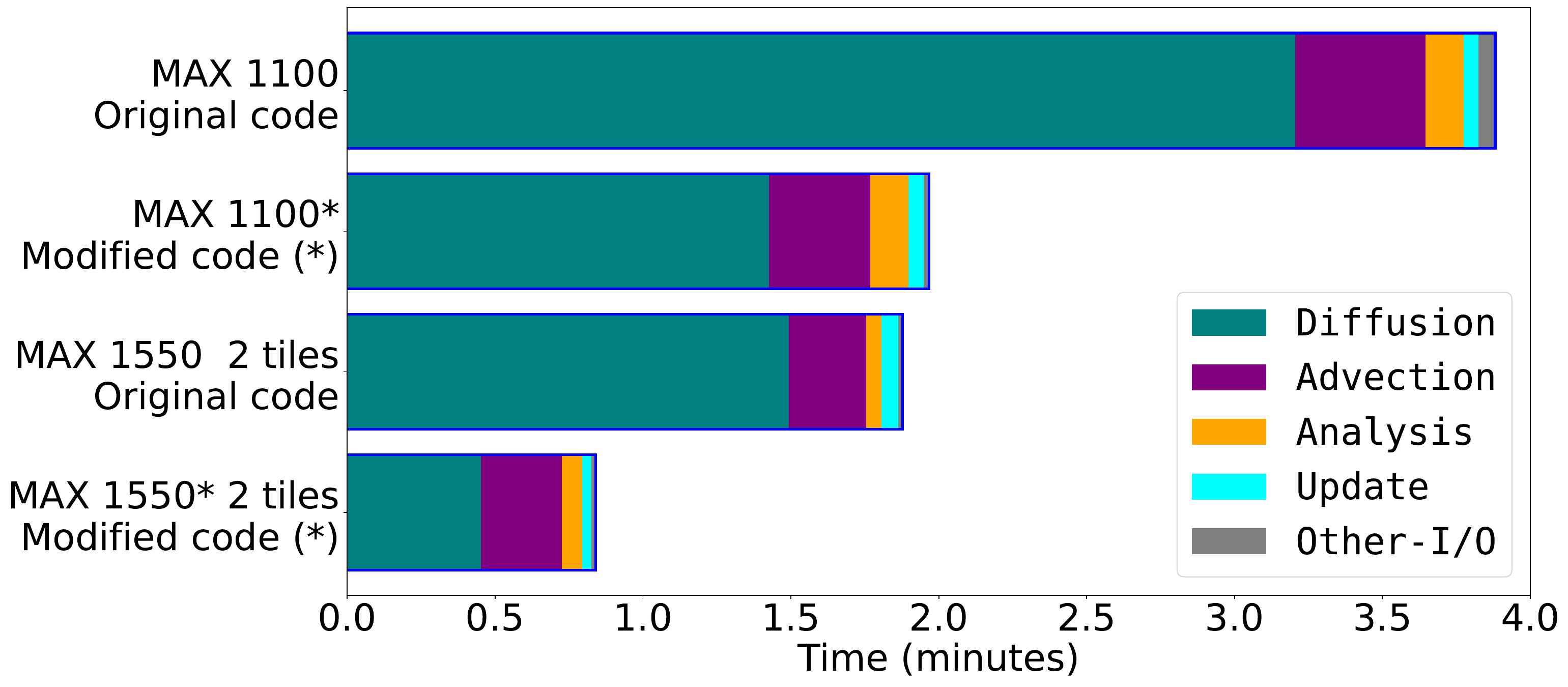}
\caption{Run time comparison (less is better) of the HipFT test case on Intel data center GPUs between the original HipFT code and the modified code (adding {\tt !\$omp parallel loop} to the inner nested DC loops).  We see a substantial performance improvement with the modified code (for these results, the file I/O time has been omitted from the "Other" category).
\label{fig:intel_fix}}
\end{figure}

As mentioned in Sec.~\ref{sec:build_intel}, the current (as of this writing 2024.2) Intel compiler does not efficiently parallelize some nested DC loops.  The problem has been identified and planned to be remedied.  In the meantime, by adding {\tt !\$omp parallel loop} to the inner nested DC loops, the current compiler efficiently parallelizes the loops.  We denote this small (6-line) modification to the code with an asterisk (*).  In Fig.~\ref{fig:intel_fix}, we show how this modification improves the performance on MAX 1100 and MAX 1550 GPUs.  The MAX 1550 GPU is designed with two tiles which can be set up to appear as a single GPU (known as "composite" mode) or as two separate GPUs ("flat" mode)\footnote{\url{https://spec.oneapi.io/level-zero/latest/core/PROG.html\#device-hierarchy}}. For all results in this paper, we use the full GPU with the 2 tiles seen as separate GPUs (1 MPI rank each) as we found the performance is better than using them together as one device.

For NVIDIA GPUs, we compare using manual memory management with OpenMP target data directives with a pure Fortran approach using automatic memory management with {\tt managed} and, for the GH, {\tt unified}.  Since the NVIDIA compiler uses data directives as data optimization hints when using {\tt unified}, to get a `pure Fortran' result, we disable this feature for our {\tt unified} runs.

\begin{figure*}[tbp]
\centering
\includegraphics[width=0.975\textwidth]{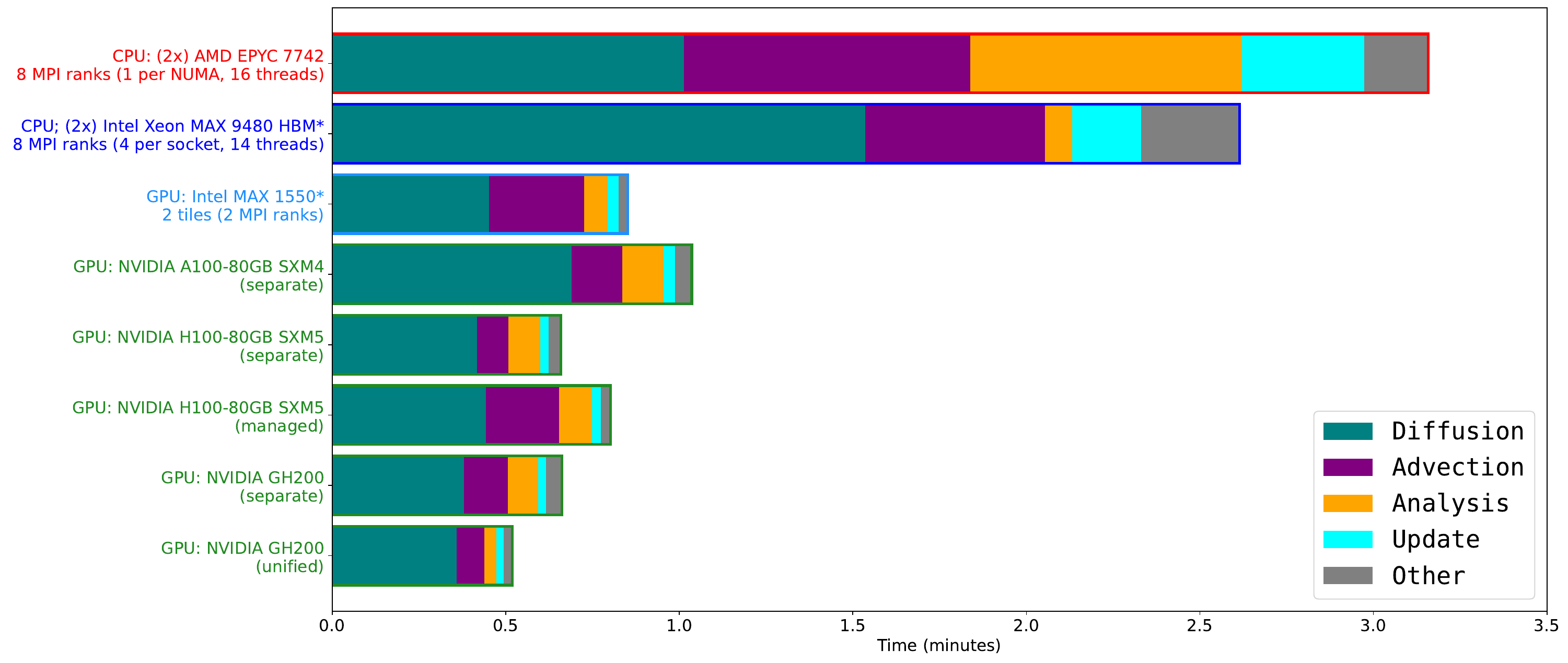}
\caption{Timing results (less is better) of the HipFT test case for server/data center CPUs and GPUs.
\label{fig:results_server}} 
\end{figure*}

The results for NVIDIA and Intel server/data center GPUs are shown in Fig.~\ref{fig:results_server} along with some CPU timings for reference.
The key takeaway is that the NVIDIA and Intel GPU runs are similar to each other considering the respective memory bandwidths of the GPUs (see the roofline plots in Fig.~\ref{fig:roofserver}), showing the increasing portability of using DC (we discuss AMD results below).  

Although we had to (temporarily) slightly modify the code to get good performance on the Intel GPUs and use OpenMP target for data movement, we were still able to use standard Fortran for the vast majority of the code.  The performance of the Intel MAX 1550 GPU is quite good considering previous difficulties in achieving maximum performance \cite{Salvadore2024}.  This may be due to running HipFT with 2 MPI ranks, one per GPU tile.  Using more MPI ranks typically improves performance on HipFT by spreading the problem out more evenly (for example, to yield best performance on the CPUs, we use the maximum number of MPI ranks possible spread evenly across NUMA domains).  We note that we did not test using multiple MPI ranks within a single NVIDIA GPU using MPS\footnote{\url{https://docs.nvidia.com/deploy/mps}} or MIG\footnote{\url{https://www.nvidia.com/en-us/technologies/multi-instance-gpu}}. This is left for future testing.

For the NVIDIA platform, we see results using automatic managed memory similar to those with manual data management.  In those cases, the code ignores the data movement directives, making the code fully standard Fortran (with the exception of a single directive for device selection).  For the NVIDIA H100, manually managed memory yields a faster result than using automatically {\tt managed} memory (consistent with results from \cite{stdpar_dc_waccpd} and \cite{caplan2023acceleration}).  However, on the GH platform, using {\tt unified} memory actually performs better than manual memory management, making the pure Fortran code not only an ideal programming method, but, in this case, the most optimal one as well.

For AMD, we were able to successfully compile HipFT using an in-development version of the CCE compiler.  The code passed the testsuite running on an MI300A APU (which has its CPU and GPU on the same chip).  While this APU has been shown to yield good performance in OpenMP target codes \cite{mi300a}, here, the run time for the test case was much slower than expected.  As the support for DC GPU offload is extremely new in CCE for AMD GPUs, we expect the performance to improve with further compiler development.  We therefore do not include the result in Fig.~\ref{fig:results_server}.  However, the ability to compile and run HipFT in pure Fortran mode correctly on an AMD GPU is a great step forward in portability.

\subsection{Consumer}
\label{sec:results_consumer}

As mentioned above, consumer GPUs, although not specialized for running HPC applications, can still be useful for development and running small (from an HPC perspective) workloads inexpensively. NVIDIA, Intel, and AMD all support running computations on consumer GPUs, but their double precision FP units are greatly reduced (or non-existent).

\begin{figure}[tb]
\centering
\includegraphics[width=0.475\textwidth]{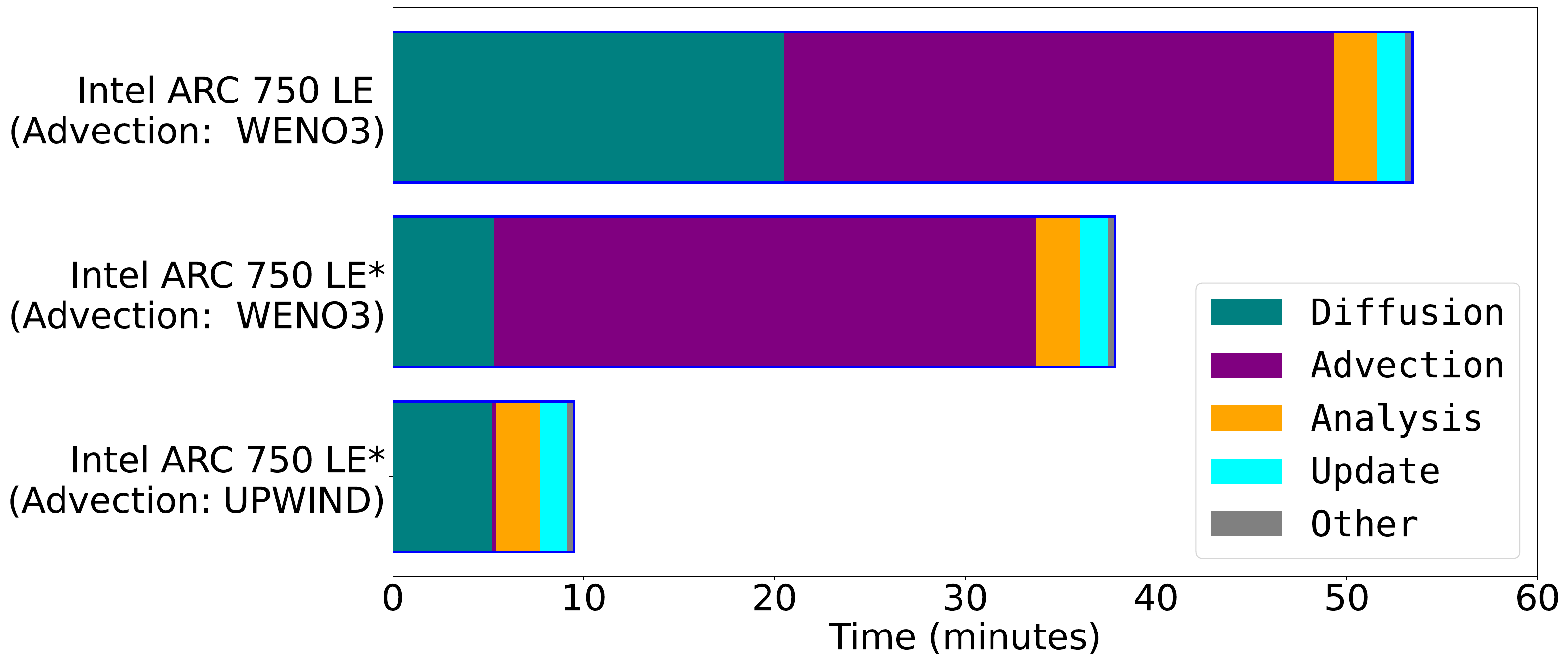}
\caption{Run times (less is better) of the HipFT test case on the Intel Arc 750 LE consumer GPU.  We show the original code, the slightly modified code (*), and the modified code running the alternative advection algorithm (upwind).
\label{fig:arc}}
\end{figure}

Looking at the roofline results for consumer hardware shown in Fig.~\ref{fig:roofconsumer} we see that they mostly do not reach their FP performance saturation within the arithmetic intensity range of HipFT's kernels, a key fact allowing their use in memory bound HPC applications.  However, due to the use of double precision emulation on the Intel Arc GPU, it does reach its FP roof and therefore is not expected to perform at the level based solely on its memory bandwidth.  It also has a wider range of bandwidth between reads and writes compared to other GPUs.  The double precision emulation turns out to be extremely slow for divisions, making the advection algorithm in the test case very slow (as the WENO3 scheme has several unavoidable division operators).  The benchmark used in Fig.~\ref{fig:roofconsumer} tests FP  performance with fused multiply-add instructions (no divisions), so the roofline shown does not reflect this slowdown.  HipFT also includes a simpler `upwind' advection scheme that has no division operators (activated by inserting the line: {\tt flow\_num\_method=1}, into the {\tt hipft.in} input file). In Fig~\ref{fig:arc} we show the test case run time on an Intel Arc 750 LE GPU for the original code, the (*) modified code, and the (*) modified code using the upwind scheme for advection.
We see that the upwind advection is many times faster than the WENO3 scheme due to the absence of divisions (on other GPUs, the upwind scheme is typically $10$ to $30$ times faster, not the $120$ times we see here).  As future Intel consumer GPUs may have hardware double precision compute units, in order to have a more representative comparison, we run the test case with the upwind method exclusively across all consumer platforms.  Note that this makes these consumer results \emph{not} comparable to the server results in Sec.~\ref{sec:results_server}. 

\begin{figure*}[tbp]
\centering
\includegraphics[width=0.975\textwidth]{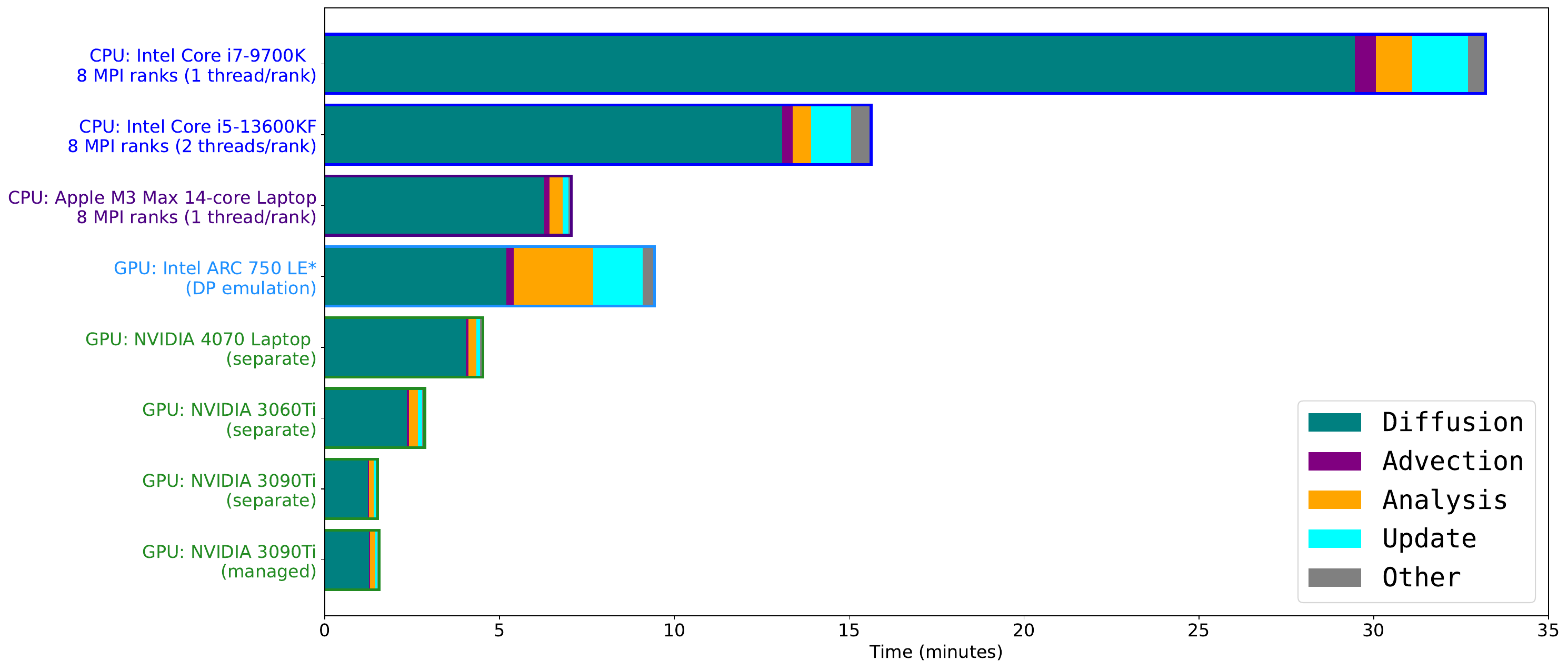}
\caption{Timing results (less is better) of the HipFT test case for consumer CPUs and GPUs.  These results use the upwind method for advection; therefore, they cannot be directly compared to the server results in Fig.~\ref{fig:results_server}.
\label{fig:results_consumer}} 
\end{figure*}

We show the timing results for the test case with the upwind scheme on consumer GPUs (and some consumer CPUs for comparison) in Fig.~\ref{fig:results_consumer}.

We see that the Intel Arc GPU does not perform at the expected level based on its memory bandwidth as expected, most likely 
 due to the double precision emulation.  However, it is still 
 capable of reasonable run times for the test case (given the upwind algorithm), outperforming many modern consumer CPUs.

For NVIDIA, we once again find that the performance using automatic {\tt managed} memory is similar to that of using manual OpenMP target data movement directives.  Thus, the consumer NVIDIA GPUs (even those on laptops) can run pure standard Fortran quite well, with performance nearing that of some server GPUs, making them a good option for development and small workloads.

The key takeaway is that the NVIDIA and Intel GPU runs are similar to each other considering the respective memory bandwidths of the GPUs (see Fig.~\ref{fig:roofconsumer}), showing the increasing portability of using DC.  

For AMD, since the current CCE DC offload implementation requires a GPU capable of unified memory, we were not able to run the test case on any consumer AMD GPUs we had access to.  However, as the Flang compiler is developing DC offload support for discrete AMD GPUs, running DC codes on consumer AMD GPUs should be possible in the future.

\section{Summary and outlook}
\label{sec:summary}

Using standard language parallelism for GPU-acceleration across multiple GPU vendors is a valuable goal for high-performance portable codes, allowing domain scientists to develop them without needing to learn to use unfamiliar external APIs.  In this paper, we have used a production Solar surface flux transport code called HipFT to test the current state of performance portability of Fortran's {\tt do concurrent} across three major GPU vendors: NVIDIA, Intel, and AMD.

Three concepts that the Fortran language is lacking that are needed/wanted for GPU computing are the recognition and management of separate memory spaces, the recognition and selection of multiple devices, and asynchronous loop execution.  With the advent of unified GPU-CPU architectures like the NVIDIA Grace-Hopper and AMD MI300A, along with automatic memory management capabilities of compilers, the data management is taken care of.  With alternate methods of launching the code (see Ref.~\cite{caplan2023acceleration} for an example), the device selection can be done without directives as well.  For discrete GPUs, manually managing the GPU-CPU data movement can be important for performance.  Therefore, augmenting standard Fortran with data directives (either OpenMP target or OpenACC) is often preferred, and for the Intel platform is currently required for good performance.

In this paper, we were able to achieve good performance on NVIDIA and Intel GPUs (with some slight modifications) and showed that the performance difference between automatic memory management and manual data movement using OpenMP target directives on platforms that support both was minimal.  We found that using automatic memory management on discrete GPUs slightly reduced performance, but when using it on a unified CPU-GPU SoC (in this case the GH200), the performance actually slightly increased.  We were also able to compile and correctly run the code on AMD GPUs, but more development is needed for good performance.

We also tested the code on consumer GPUs where possible, and showed good performance, especially for NVIDIA GPUs.  Reasonable performance was also achieved for some algorithms on an Intel Arc GPU even though it is not designed for HPC workloads and requires the use of double precision emulation.  We expect that the performance on Intel's upcoming consumer GPUs will be much better as they may have hardware double precision units. We were not able to test performance on consumer AMD GPUs, but we expect support for those is forthcoming.

We note that HipFT is not a large code ($\sim9000$ lines) and does not use certain features that are known to be challenging for GPU acceleration (such as GPU-aware MPI, derived type arrays, and function calls within parallel regions \cite{caplan2019gpu}).  We plan to continue testing DC across vendors for larger codes that do contain those features.

We highly encourage vendors to continue to improve standard language parallelism, including the support and interoperability of directive APIs (both OpenMP and OpenACC).  These APIs can augment the standard languages with required or optimal features until such time that the language standards are extended to no longer need them.

\section*{Acknowledgment}
Work at Predictive Science Inc. was supported by the NASA LWS Strategic Capabilities Program (grant 80NSSC22K0893), the NSF PREEVENTS program (grant ICER1854790), and the NSF/NASA SWQU program (grants AGS 2028154 and 80NSSC20K1582).  It also utilized the Cabeus system at NASA's HECC through NASA grant's 80NSSC20K0192's NAS request SMD-24-72380598, as well as the Expanse system at SDSC and the Stampede3 system at TACC through ACCESS allocation TG-MCA03S014.  

We also thank Erika Palmerio, Ryder Davidson, Jeff Sandoval, Gary Elsesser, and Aaron Jarmusch for their assistance and suggestions toward completion of this work.

\section*{Appendix}
\label{sec:appendix}

\section{Roofline plots of tested hardware}
Here we use the OpenCL-Benchmark \cite{openclbench} to produce roofline plots of the GPUs tested in this paper.  The optimized and simple computations of the benchmark make it ideal as a practical (instead of peak theoretical) way to generate roofline plots for comparison.  We use the benchmark's floating point and memory read/write results to show the best FLOP/s expected given a kernel/algorithm's arithmetic intensity (FLOP/s per byte).  The floating point results in the OpenCL-Benchmark are computed with fused multiply adds and appear as
\begin{lstlisting}[language=Compile, label={lst:openclfp}]
kernel_double(global float* data){
  double x = (double)get_global_id(0);
  double y = (double)get_local_id(0);
  for(uint i=0u;i<128u;i++){
    x = fma(y, x, y);
    y = fma(x, y, x);}
  data[get_global_id(0)] = (float)y;}
\end{lstlisting}
while the coalesced read and write kernels are computed as
\begin{lstlisting}[language=Compile, label={lst:openclfp}]
kernel_coalesced_write(global float* data){
  const uint n = get_global_id(0);
  for(uint i=0u;i<def_M;i++) data[i*def_N+n]=0.0f;}
kernel_coalesced_read(global float* data){
  const uint n = get_global_id(0);
  float x = 0.0f;
  for(uint i=0u;i<def_M;i++) x += data[i*def_N+n];
  data[n] = x;}
\end{lstlisting}

\begin{figure}[htb]
\centering
\includegraphics[width=0.475\textwidth]{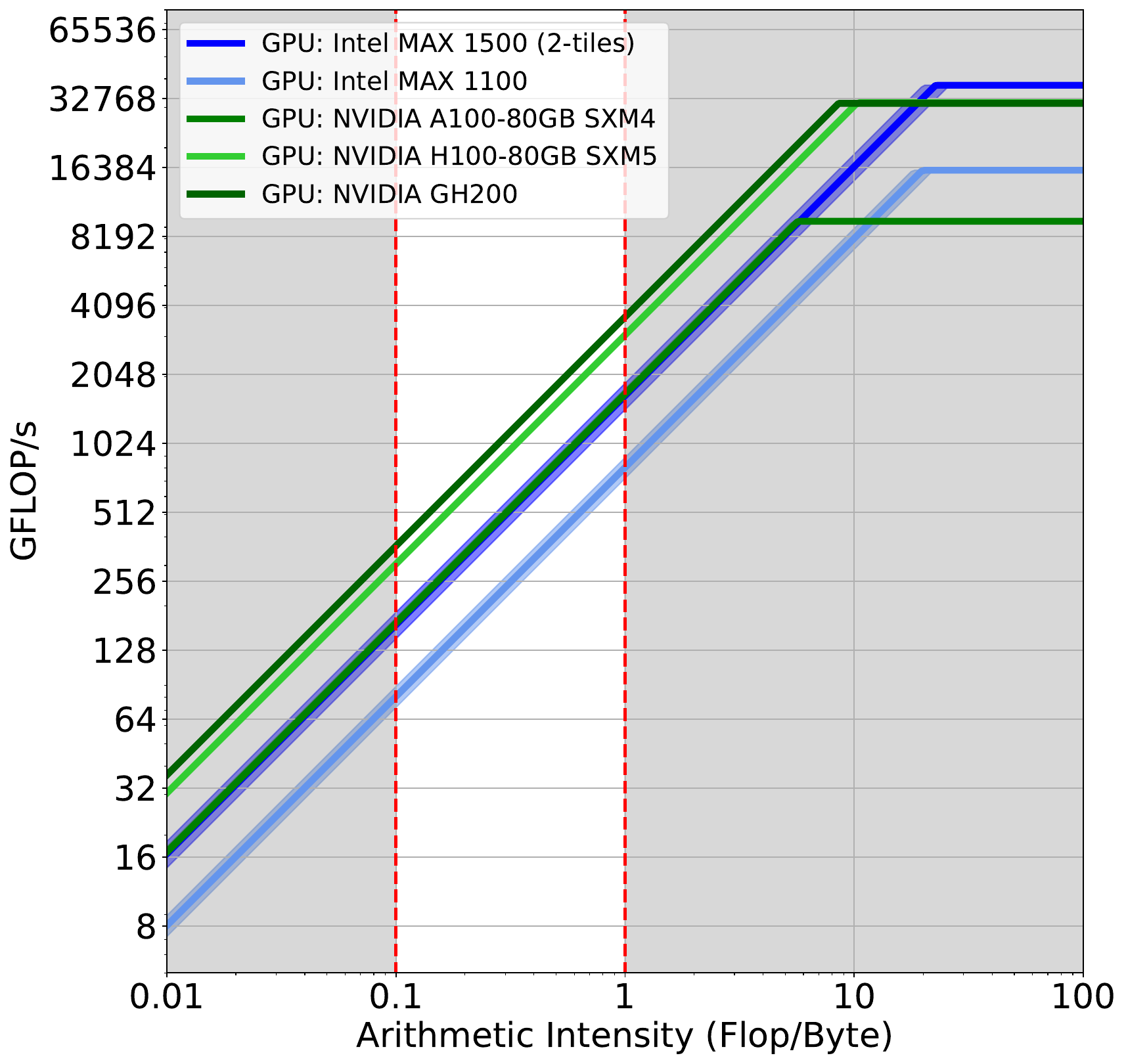}
\caption{Roofline plots of the server GPUs tested.  The double precision FLOP/s and memory bandwidth values are computed using the OpenCL-Benchmark \cite{openclbench}.  The memory bandwidth is taken as the average of coalesced reads and writes, with the shaded regions spanning the minimum and maximum across both. The range of arithmetic intensities of kernels used in HipFT is shown as the region not shaded in gray.
\label{fig:roofserver}}
\end{figure}

\begin{figure}[htb]
\centering
\includegraphics[width=0.475\textwidth]{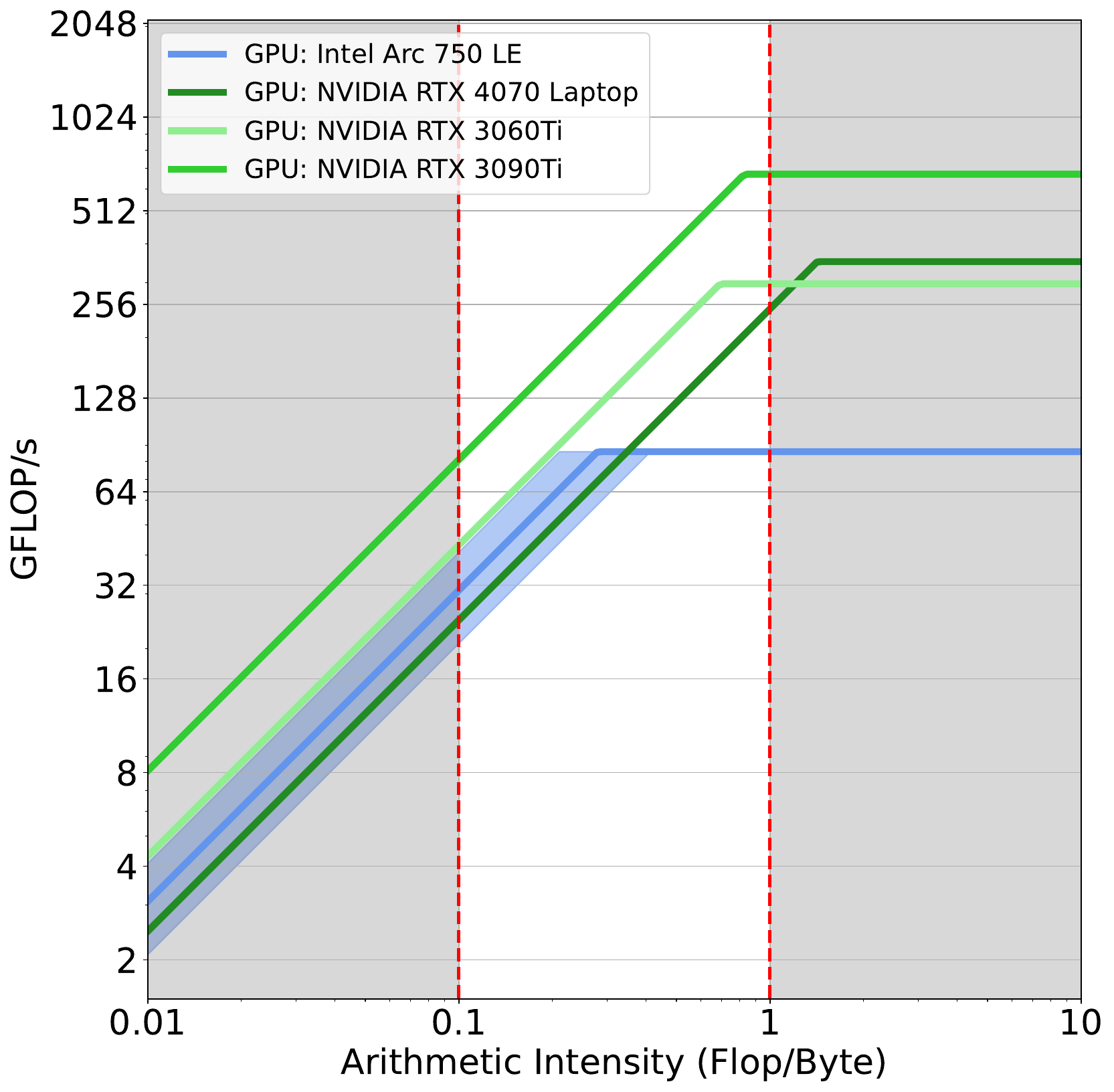}
\caption{Roofline plots (double precision) of the consumer GPUs tested, computed with the OpenCL-Benchmark \cite{openclbench}.  See Fig.~\ref{fig:roofserver} for description.
\label{fig:roofconsumer}}
\end{figure}

The server/data center results are shown in Fig.~\ref{fig:roofserver}, while those for consumer hardware are shown in Fig.~\ref{fig:roofconsumer}. In each, the range of arithmetic intensities used in HipFT is highlighted.

\section{Artifact Availability Statement}

\newcommand*\ttvar[1]{\texttt{\expandafter\dottvar\detokenize{#1}\relax}}
\newcommand*\dottvar[1]{\ifx\relax#1\else
  \expandafter\ifx\string/#1\string/\allowbreak\else#1\fi
  \expandafter\dottvar\fi}

\section*{Artifact Identification}
We have investigated the portability of using Fortran's `do concurrent` (DC) for accelerated computing by testing its use across GPU vendors using the solar flux evolution code HipFT.  HipFT is written using DC loops and includes OpenMP target data directives for manual data movement between the CPU and GPU for platforms that do not support (or have non-optimal) automatic data management.  We successfully compiled and ran a production-level test case on NVIDIA, Intel, and AMD GPUs using NVIDIA's HPC SDK's {\tt nvfortran}, Intel's OneAPI HPC Toolkit's {\tt ifx}, and HPE's Cray Compiler Environment's {\tt ftn} compilers respectively. Both server and consumer hardware were tested.  Tags/releases of the github repository of HipFT (\url{https://github.com/predsci/hipft}) were made for the versions used in this paper, specifically the {\tt v1.11.0} release of the code was used for CPU runs and the NVIDIA GPU and AMD GPU runs. The {\tt v1.11.0-intel-gpu} branch release (only 6 modified lines) was used for Intel GPU runs.    The test case is the run located in the {\tt examples/flux\_transport\_1rot\_flowAa\_diff\_r8} folder of the HipFT github repository.  Given access to the computational hardware, the repository with its release tags can be used to directly reproduce the run timings shown in the paper by building with the compiler flags described in the paper, along with any/all system-specific compiler flags that may be needed.  The HipFT code outputs a breakdown of the time taken in each section of the code, and this data is used to generate the bar plots in the figures.

\section*{Reproducibility of Experiments}
The workflow to generate the results in this paper is as follows:  First, checkout the relevant release of HipFT (either \url{github.com/predsci/HipFT/releases/tag/v1.11.0} for NVIDIA and AMD GPUs or \url{github.com/predsci/HipFT/releases/tag/v1.11.0-intel-gpu} for Intel GPUs).  Next, build the code according to the build instructions (including having an HDF5 library installed with the same compiler used for HipFT), the included sample build scripts, and the compiler flags described in the paper.  Then, copy the input files from the {\tt examples/flux\_transport\_1rot\_flowAa\_diff\_r8} folder into a new folder, and run the code with {\tt mpirun -np <N> hipft} (assuming HipFT is in your run path) and where {\tt <N>} is the number of MPI ranks to use (in this case, typically 1).

The execution time will depend on the hardware and software platforms.  If running on similar hardware as described in the paper, the run times should be very close to  those in the bar plots (maximum $\sim1$hr), except possibly for the gray "other" area which can be sensitive to the speed of the file system and network.

If the test case is run on the same hardware and software versions (including compilers) as done in the paper, it is expected to yield very similar run times as those shown in the bar plots.  The results should show the portability of Fortran's DC for GPU acceleration, with similar timings for NVIDIA and Intel GPUs of comparable capabilities, as well as correct execution on AMD GPUs.   Running the tests with newer versions of the compilers, or on newer GPU hardware should result in faster run times.

%
%
%
\bibliographystyle{IEEEtran}
\bibliography{ref.bib}

\end{document}